\documentclass{article}
\usepackage{spconf,amsmath,graphicx,hyperref} 
\usepackage{float}

\title{Tracking Listener Attention: Gaze-Guided Audio-Visual Speech Enhancement Framework}

\name{
Hsiang-Cheng Yang$^{*1}$,
You-Jin Li$^{*12}$,
Rong Chao$^{12}$,
Yu Tsao$^{2}$,
Borching Su$^{1}$,
Shao-Yi Chien$^{1}$
\thanks{*These authors contributed equally to this work.}
}
\address{
$^{1}$ National Taiwan University, Taipei, Taiwan \\
$^{2}$ Academia Sinica, Taipei, Taiwan \\
\texttt{r12943011@ntu.edu.tw},
\texttt{yu.tsao@citi.sinica.edu.tw}
}

\begin{document}
\ninept
\maketitle

\begin{abstract}

This paper presents a Gaze-Guided Audio-Visual Speech Enhancement (GG-AVSE) framework to address the cocktail party problem. A major challenge in conventional AVSE is identifying the listener’s intended speaker in multi-talker environments. GG-AVSE addresses this issue by exploiting gaze direction as a supervisory cue for target-speaker selection. Specifically, we propose the GG-VM module, which combines gaze signals with a YOLO5Face detector to extract the target speaker’s facial features and integrates them with the pretrained AVSEMamba model through two strategies: zero-shot merging and partial visual fine-tuning. For evaluation, we introduce the AVSEC2-Gaze dataset. Experimental results show that GG-AVSE achieves substantial performance gains over gaze-free baselines: a 10.08\% improvement in PESQ (2.370 → 2.609), a 5.18\% improvement in STOI (0.8802 → 0.9258), and a 23.69\% improvement in SI-SDR (9.16 → 11.33). These results confirm that gaze provides an effective cue for resolving target-speaker ambiguity and highlight the scalability of GG-AVSE for real-world applications.

\end{abstract}


\section{Introduction}
\label{sec:intro}
The cocktail party problem~\cite{cherry1953some} refers to the challenge of isolating a target speaker’s voice in noisy, multi-speaker environments. This issue is particularly critical for applications such as hearing assistive technologies~\cite{ha1,ha2}, smart cockpits, and video conferencing systems. Despite substantial progress, traditional audio-only enhancement methods continue to struggle in multi-speaker scenarios where the target voice cannot be reliably separated from competing sources. In contrast, the human auditory system performs remarkably well in such complex conditions by integrating both auditory and visual cues. When audio signals are masked or degraded by noise, visual information, such as lip movements and facial expressions, remains robust and provides reliable supplementary cues. Motivated by this multisensory capability, the primary goal of Audio-Visual Speech Enhancement (AVSE)~\cite{avse1,avse2,avse3,avse4} is to fuse complementary auditory and visual information to reconstruct clear and intelligible speech even under highly adverse acoustic conditions.



In recent years, advances in deep learning have driven rapid progress in AVSE~\cite{michelsanti2021overview}. Early studies typically concatenated audio and visual features and employed architectures such as Convolutional Neural Networks (CNNs) or U-Nets, with visual input largely restricted to the speaker’s lip movements. To better address real-world challenges such as facial occlusions and pose variations, later research emphasized improving model robustness. At the same time, the scope of visual input expanded from the mouth region to the full face, and Self-Supervised Learning (SSL)~\cite{10193049} methods were introduced to enhance cross-modal alignment and consistency. More recent work has explored advanced architectures such as Conformers, which leverage scene-level visual cues beyond the face to further improve separation quality. To address scalability limitations in processing long sequences and to reduce model complexity, AVSEMamba~\cite{chao2025leveraging} has been proposed as a hybrid AVSE framework. It integrates full-face video and audio features through a Mamba-based temporal–frequency model, offering both computational efficiency and favorable memory scaling~\cite{gu2021efficiently}.


Despite the strong performance and demonstrated potential of AVSE technology, existing systems face a critical bottleneck for practical deployment: reliably extracting the correct visual features during inference. In single-speaker video frames, public face extraction models such as YOLO~\cite{redmon2016you}, RetinaFace~\cite{deng2020retinaface}, and MTCNN~\cite{zhang2016joint} can capture facial features with high accuracy. However, in multi-speaker scenarios, these models cannot determine which individual corresponds to the listener’s intended target, thereby limiting the AVSE system’s ability to perform effective enhancement.


This paper proposes a novel Gaze-Guided Visual Module (GG-VM) framework that leverages the listener’s gaze to identify their visual focus. By integrating gaze information with the YOLO5Face model~\cite{qi2022yolo5face}, the framework dynamically captures the facial features of the intended speaker, enabling speech enhancement (SE) that can seamlessly switch targets based on the listener’s gaze. In our experiments, we evaluate the effectiveness of this framework within the AVSE paradigm, comparing zero-shot and fine-tuned visual models. Results show that both approaches reliably capture target speaker features, thereby facilitating accurate target speech extraction. Overall, the proposed technology represents an important step toward real-world deployment of AVSE in applications such as wearable devices, smart glasses, and Mixed Reality (MR).

\section{Related Work}
\label{sec:related}

\subsection{Mamba-based audio-visual speech enhancement}
The primary objective of a Speech Enhancement (SE) system is to recover a clean target signal $s(t)$ from a noisy observation $y(t)$, which is typically modeled as:
\begin{equation}
y(t) = s(t) + v(t) + n(t),
\label{eq:se_model}
\end{equation}
where $v(t)$ and $n(t)$ represent interfering speech and background noise, respectively. While single-channel audio-only SE has made significant strides, it often struggles in low-SNR conditions or "cocktail party" scenarios where acoustic characteristics alone are insufficient to separate overlapping sources~\cite{cherry1953some}.

To overcome these limitations, AVSE integrates a synchronized visual stream $x_{v}(t)$—typically comprising lip or facial movements—into the enhancement process. The estimation can be formulated as:
\begin{equation}
\hat{s}(t) = f_{\theta}(y(t), x_{v}(t)).
\label{eq:avse_model}
\end{equation}
Since visual cues are immune to acoustic noise, they provide robust guidance for speech reconstruction.

In terms of architecture, the field has evolved from CNNs and U-Nets to more advanced attention-based models like Transformers and Conformers. However, the self-attention mechanism in Transformers suffers from quadratic computational complexity ($O(L^2)$) with respect to sequence length $L$, making it computationally expensive for long audio sequences. To address this efficiency bottleneck, Structured State Space Models (SSMs)~\cite{gu2021efficiently} have emerged as a compelling alternative. Specifically, the Mamba architecture~\cite{gu2023mamba} introduces a selective state space mechanism that achieves linear-time complexity ($O(L)$) while maintaining the ability to model long-range temporal dependencies. Leveraging this, AVSEMamba~\cite{chao2025leveraging} was proposed as a hybrid framework that fuses full-face visual features with audio spectrograms using Time-Frequency Mamba blocks, offering a superior trade-off between inference speed and enhancement quality.

\subsection{Toward gaze-aware multimodal enhancement}
Despite the architectural advancements, a fundamental limitation persists in most AVSE systems: the assumption that the visual input always corresponds to the desired target speaker. In realistic multi-talker environments, multiple faces are often visible simultaneously, creating ambiguity that standard models cannot resolve.

This challenge is particularly critical for human-centric applications such as hearing assistive technologies, where the system must dynamically align with the user's intent. Recent research has begun to explore auxiliary cues to bridge this gap. For instance, Padilla et al.~\cite{padilla2025location} proposed a location-aware target speaker extraction method for hearing aids, demonstrating that spatial information can effectively filter out interference in complex acoustic scenes. In the visual domain, Anway et al.~\cite{anway2024real} introduced a real-time gaze-directed speech enhancement framework, validating that eye-tracking signals can serve as a natural and intuitive interface for selecting the attended speaker in audio-visual hearing aids.

However, existing gaze-guided approaches often rely on heavier computational backbones or specific hardware constraints. Our proposed GG-AVSE framework builds upon these user-centric paradigms but distinguishes itself by integrating explicit gaze guidance directly with the lightweight AVSEMamba architecture. By combining the efficiency of Mamba with a modular Gaze-Guided Visual Module (GG-VM), our approach enables robust, low-latency target extraction that is scalable to wearable devices, effectively resolving the target ambiguity problem in multi-person video scenes.

\section{Proposed Method}
\label{sec:Proposed}
In this study, we propose the Gaze-Guided Audio-Visual Speech Enhancement (GG-AVSE) framework, which comprises two key components: a GG-VM and an AVSEMamba model with visual encoder fine-tuning.

\begin{figure}[]
    \centering
    \includegraphics[width=0.8\linewidth]{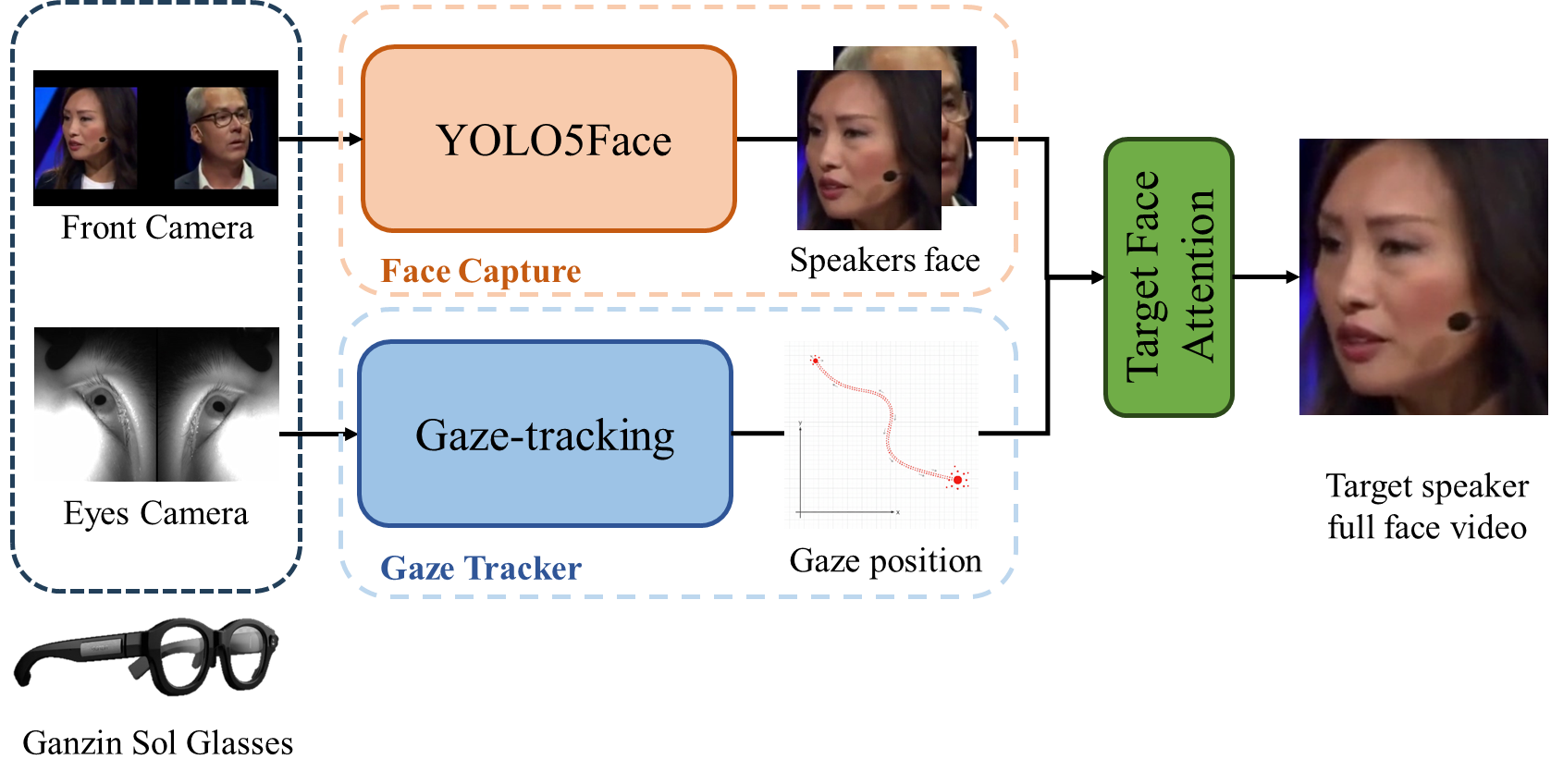}
    \caption{System architecture of the proposed GG-VM module.}
    \label{fig:GGextorctor}
\end{figure}

\subsection{Gaze-guided visual module}
\label{ssec:ggffe}
Identifying the attended speaker is essential in multi-speaker scenarios. We address this challenge by tracking the listener’s visual attention through the GG-VM module, illustrated in Figure~\ref{fig:GGextorctor}. GG-VM consists of three components: a gaze tracker, a face capture module, and a target face attention system. To realize GG-VM, we employed the Ganzin Sol Glasses~\cite{GSG}, a wearable eye-tracking device that provides frontend visual and gaze data.

\textbf{Gaze Tracker:} The Ganzin Sol Glasses employ image processing and machine learning techniques to extract the user’s eye position and gaze point. They provide accurate real-time tracking at a 120 Hz sampling rate, where the high-frequency sampling ensures stable gaze estimation and reliable acquisition of the user’s current gaze point, which we use as a guiding signal to identify the target speaker.

\textbf{Face Capture:}
YOLO5Face~\cite{qi2022yolo5face} is specifically designed for face detection and incorporates face-specific anchor priors, which improve robustness on small and partially occluded faces, compared to generic object detectors. On the WIDER FACE benchmark~\cite{yang2016wider}, YOLO5Face achieves a favorable balance between accuracy and inference speed. In comparison, RetinaFace~\cite{deng2020retinaface} offers higher accuracy but with significantly greater latency, while MTCNN~\cite{zhang2016joint} is faster but exhibits poor recall on small faces. For our AVSE system, which demands both robustness and low latency, YOLO5Face provides the most suitable trade-off.

\textbf{Target Face Attention:}
To associate gaze with detected faces, we designed a matching score that incorporates both spatial distance and region overlap:
\begin{equation}
Score(i) = \gamma \cdot D(i) + (1 - \gamma) \cdot IoU(i), \quad \gamma \in [0,1]
\end{equation}
where $D(i)$ denotes the inverse Euclidean distance between the gaze point and the face center, while $IoU(i)$ measures the overlap between the gaze region and the detected face box~\cite{everingham2010pascal}. A larger $IoU$ is desirable when multiple faces are close together, whereas $D$ is more reliable under normal spacing. The weight $\gamma$ balances these two cues and is empirically set to 0.75 to maximize correct speaker recognition on the validation set.



The proposed matching score effectively combines the strengths of geometric proximity
and region-level overlap, enabling robust gaze-to-face association without relying
on explicit identity tracking. Similar design principles have been successfully
adopted in object detection, such as the Distance-IoU loss~\cite{zheng2020distance},
which demonstrates the benefit of jointly considering distance and overlap.
Inspired by this principle, our formulation maintains both interpretability and
computational efficiency, making it well suited for real-time AVSE applications.

\begin{figure*}[]
    \centering
    \includegraphics[width=0.9\textwidth]{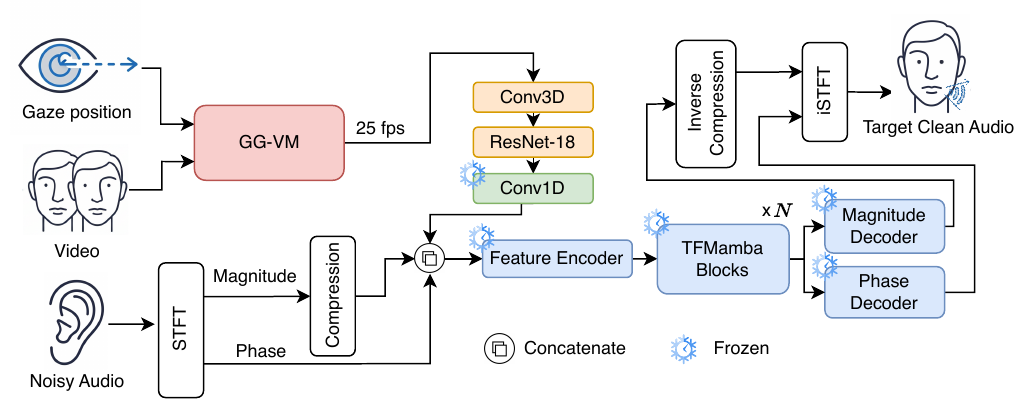}
    \caption{System architecture of the proposed GG-AVSE model.}
    \label{fig:avsemamba}
\end{figure*}

\subsection{The proposed GG-AVSE framework}
To demonstrate the effectiveness of the proposed GG-VM module, we employ a SOTA
pretrained AVSE model rather than training from scratch. Specifically, we adopt AVSEMamba—the top-performing system in AVSEC-4—as our base model. The resulting GG-AVSE framework integrates GG-VM with AVSEMamba using two strategies: zero-shot merging and partial visual fine-tuning (PVFT), as illustrated in Figure~\ref{fig:avsemamba}.

\textbf{Zero-Shot Merge:} This method is the most cost-effective implementation,
as it requires no model retraining and operates under a single-channel audio setting.
The process involves two key preprocessing steps: (1) calibrating the output of the
GG-VM module to match the dimensionality of AVSEMamba’s visual encoder input, and
(2) normalizing the scale of full-facial visuals from GG-VM to ensure consistency
with the face sizes in the original pretraining dataset. Experimental results show that with these adjustments, GG-VM integrates seamlessly with AVSEMamba. Notably, the original visual frontend was trained on single-speaker lip-reading data; yet, despite this domain gap, the pretrained model yields reliable visual embeddings in our multi-speaker, gaze-guided scenario, demonstrating strong zero-shot capability.

\textbf{PVFT:} To address the visual domain discrepancy between the data captured by
Ganzin Sol Glasses and the images in the original pretraining dataset, we employ
the PVFT strategy for the visual encoder. Specifically, we adopt the SimCLR~\cite{chen2020simple}
contrastive learning framework, where positive pairs are formed from different
augmented views of the same video clip, and negative pairs are sampled from other video clips within the same mini-batch. The objective of this fine-tuning is to align the feature space, ensuring that the embeddings produced by the visual encoder for GG-VM
data are consistent with those from the pretraining data. By bridging this domain gap, the GG-AVSE architecture can more robustly extract features of the target speaker, ultimately leading to improved SE performance. This demonstrates not only the
effectiveness of our approach but also its scalability to practical AVSE deployments.

\section{Experiment}
\label{sec:exp}

To evaluate the proposed framework, we conduct comprehensive experiments on a newly constructed dataset, AVSEC2-Gaze.

\subsection{The AVSEC2-Gaze dataset}
\label{sec:subdata}


The AVSEC2-Gaze dataset was constructed as a set of gaze-guided two-speaker mixtures derived from the AVSE Challenge dataset (AVSEC-2)~\cite{10023284}. Clean speech signals were sourced from the Lip Reading Sentences 3 (LRS3) dataset~\cite{afouras2018lrs3}, while noisy signals were prepared using three noise corpora: the Clarity Enhancement Challenge (CEC1)~\cite{graetzer2021clarity}, DEMAND~\cite{thiemann2013diverse}, and the second Deep Noise Suppression (DNS) Challenge~\cite{reddy2021icassp}.

To facilitate mixing, the duration distribution of the original LRS3 dataset was first analyzed, and speech segments of similar lengths were paired. This strategy reduces distortion caused by trimming or zero-padding and enables straightforward, reproducible alignment. Target–interferer pairs were then generated exclusively from different speakers, with each utterance used once as the target and once as the interferer to further augment the dataset.

For mixing, the interferer segment was trimmed or padded to match the target length. The two signals were then combined with equal gain, and the resulting mixture was clipped to [-1, 1] to prevent overflow:

\begin{equation}
\tilde{x}_{\text{int}}[n] = \text{clip}(x_{\text{int}}[n]), \quad 0 \leq n < L_{\text{tgt}}.
\end{equation}

\begin{equation}
y[n] = \text{clip}(x_{\text{tgt}}[n] + \tilde{x}_{\text{int}}[n], -1, 1), \quad 0 \leq n < L_{\text{tgt}}.
\end{equation}

As a final step, we constructed the AVSEC2-Gaze dataset by combining target and interferer speech segments into side-by-side videos to simulate multi-speaker scenarios. 
In each frame, the attended speaker was determined by the recorded gaze coordinates from human participants, mapped to the corresponding face bounding boxes.
The dataset consists of 1000 training samples for model fine-tuning and 200 testing samples for standard evaluation. To prevent data leakage, the test sets were recorded independently from the original LRS3 test set, and an automated script was used to verify audio–visual synchronization, ensuring a consistent frame rate of 25 FPS and an audio sampling rate of 16 kHz.
\subsection{Experimental results}
\label{ssec:subperf}

\textbf{Evaluation Metrics}:
We evaluated enhancement performance using three widely adopted metrics. PESQ~\cite{rix2001perceptual} measures the perceptual quality of the enhanced speech, STOI~\cite{taal2010short, taal2011algorithm} assesses intelligibility, and SI-SDR~\cite{le2019sdr} (in dB) evaluates separation quality in decibels. For all three metrics, higher values indicate better SE performance.

\textbf{Quantitative Results}:
To evaluate the effectiveness of the proposed GG-AVSE method compared with AVSEMamba(\textit{AVSE}), we report results on the AVSEC-2~\cite{10023284}, AVSEC-2$_{200}$, and AVSEC2-Gaze datasets, as summarized in Table~\ref{tab:all_results}. 
AVSEC-2$_{200}$ is constructed by selecting samples from AVSEC-2 that match the same subjects and time intervals as AVSEC2-Gaze, forming a fixed-target test set without eyeglass-perspective recordings. \textit{Noisy} denotes the original unenhanced speech. While AVSE performs well on AVSEC-2$_{200}$, its performance degrades under more challenging multi-speaker conditions. Providing visual features from multiple simultaneous talkers without gaze filtering (\textit{AVSE$_{NL}$}) leads to a notable performance drop, highlighting the difficulty of attended-speaker identification. Incorporating gaze information (GG-AVSE) substantially improves performance, with further gains achieved through fine-tuning (GG-AVSE$_{FT}$).

Next, Table~\ref{tab:all_results} shows that although \textit{AVSE} performs well on AVSEC-2$_{200}$ (fixed-target speaker system), our gaze-guided
framework consistently outperforms it under multi-speaker conditions. Compared to \textit{AVSE}, \textit{GG-AVSE$_{FT}$} achieves notable performance improvements of 10.08\% in PESQ (2.370 → 2.609), 5.18\% in STOI (0.8802 → 0.9258), and 23.69\% in SI-SDR (9.16 → 11.33). The results confirm that combining gaze information with lightweight fine-tuning, which improves visual feature quality,  alleviates performance degradation from speaker overlap, thereby enabling robust and practical AVSE in real-world conversational environments.


\begin{table}[h]
\vspace{-2mm}
\caption{SE performance on AVSEC-2 and AVSEC2-Gaze (Gaze).}
\label{tab:all_results}
\centering
\begin{tabular}{lcccc}
\hline
\textbf{Method} & \textbf{Evalset} & \textbf{PESQ}$\uparrow$ & \textbf{STOI}$\uparrow$ & \textbf{SI-SDR$\uparrow$} \\
\hline
Noisy & AVSEC-2 & 1.137 & 0.6180 & -5.21 \\
AVSE & AVSEC-2 & 2.178 & 0.8553 & 9.06 \\
\hline
Noisy & AVSEC-2$_{200}$  & 1.227 & 0.6578 & -2.59 \\
AVSE & AVSEC-2$_{200}$ & 2.370 & 0.8802 & 9.16 \\
\hline
AVSE$_{NL}$ & Gaze & 1.488 & 0.6318 & -3.47 \\
GG-AVSE & Gaze & 2.589 & 0.9245 & 11.09 \\
GG-AVSE$_{FT}$ & \textbf{Gaze} & \textbf{2.609} &
\textbf{0.9258} & \textbf{11.33} \\
\hline
\end{tabular}
\vspace{-2mm}
\end{table}

\begin{figure}[h]
    \centering
    \includegraphics[width=0.7\linewidth]{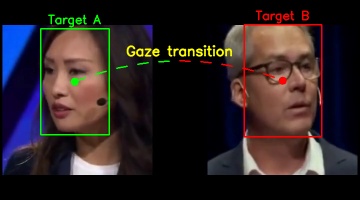}
    \includegraphics[width=1.0\linewidth]{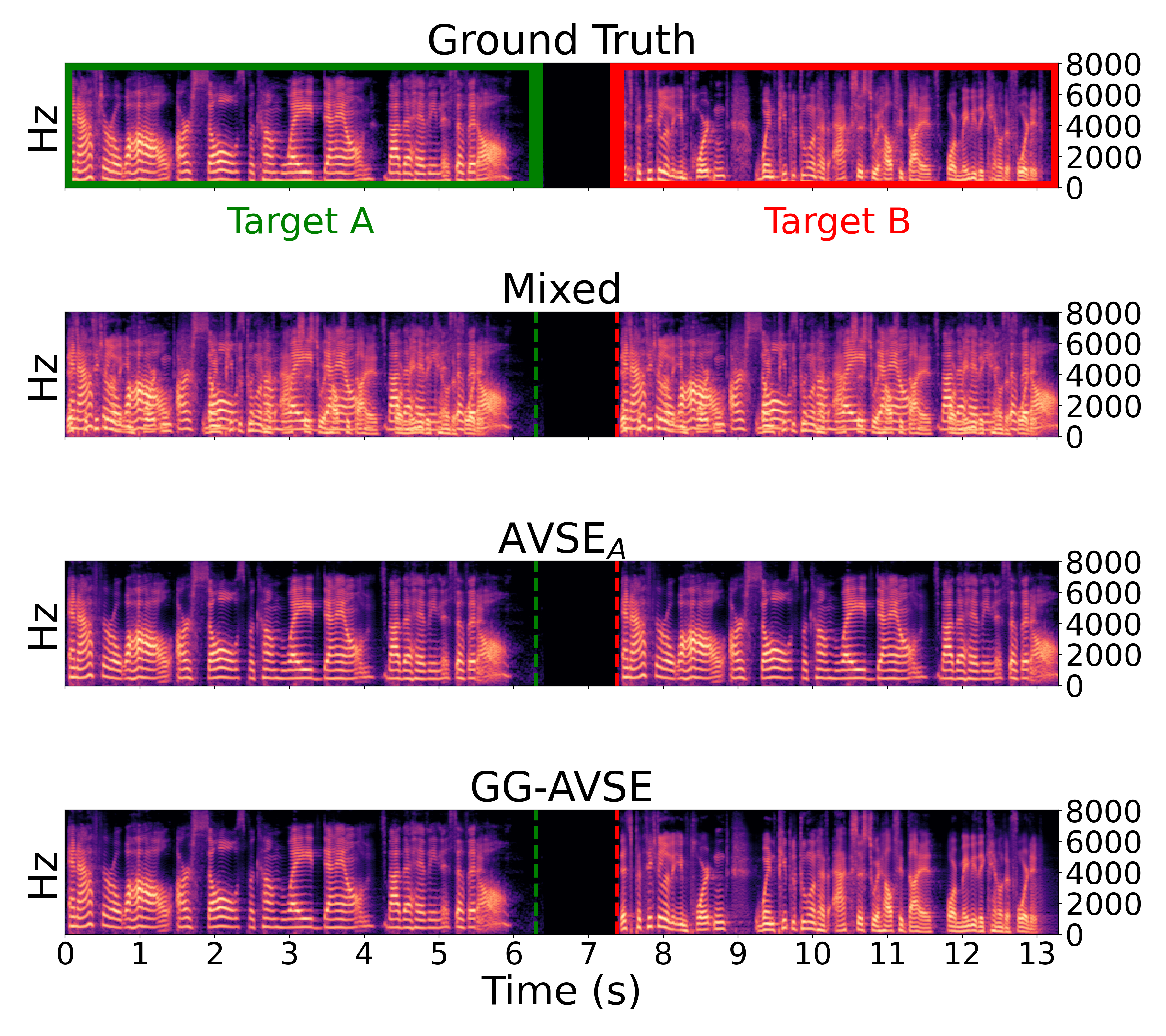}
    \caption{AVSE with two simultaneous speakers under gaze guidance. From top to bottom: \textit{Ground Truth}, \textit{Mixed, \textit{AVSE$_A$}, and \textit{GG-AVSE}. Demo link: https://jinliyou1991.github.io/GG-AVSE-Demo/}} 
    \label{fig:combined}
\end{figure}

\textbf{Tracking Target Speaker Voice in Multi-person Video Scenes:} In Figure~\ref{fig:combined}, we evaluate a scenario with a gaze transition from Target A (green box) to Target B (red box). To systematically validate this setting, we additionally constructed 100 testing samples in the AVSEC2-Gaze dataset. Each sample consists of two identical copies of the same two-speaker mixture, with a short silent gap inserted in the middle to simulate a natural gaze transition. This design provides a clear cue for the gaze shift and facilitates the alignment of ground-truth references, which are obtained by concatenating the clean speech of the attended speaker before and after the transition.

The spectrogram results are shown for four conditions: \textit{Ground Truth}, \textit{Mixed}, \textit{AVSE$_A$} (fixed-target mode, Target A), and \textit{GG-AVSE}. The \textit{Ground Truth} presents the clean reference signals, while \textit{Mixed} shows the overlapped mixture where the attended speaker is heavily corrupted by interference. In the \textit{AVSE$_A$} condition, the system remains conditioned on the initial talker and thus fails to adapt after the gaze transition, leading to suppression of the true target and degraded performance. In contrast, the proposed method dynamically follows the gaze shift and selectively enhances the corresponding target speaker, achieving clear separation before and after the transition. Table~\ref{tab:ttsvimpv} further quantifies this effect, where the GG-AVSE system consistently outperforms both the mixture and fixed-target baselines across all evaluation metrics.

\begin{table}[h]
    \centering
    \vspace{-2mm}
    \caption{SE performance under speaker-switch scenario.}
    \label{tab:ttsvimpv}
    \renewcommand{\arraystretch}{1.2}
    \vspace{9pt}
    \begin{tabular}{lccc}
    \hline
    \textbf{Method} & \textbf{PESQ} $\uparrow$ & \textbf{STOI} $\uparrow$ & \textbf{SI-SDR $\uparrow$} \\
    \hline
    Mixed & 1.238 & 0.7247 & 0.03 \\
    AVSE$_{A}$ & 1.571 & 0.7609 & 3.84 \\
    AVSE$_{B}$ & 1.602 & 0.7671 & 3.97\\
    GG-AVSE & \textbf{2.225} & \textbf{0.8380} & \textbf{7.76} \\
    \hline
\end{tabular}
\parbox{\linewidth}{\centering \footnotesize AVSE$_A$ and AVSE$_B$ denote fixed-target mode.}
\vspace{-2mm}
\end{table}

\section{Conclusion}
\label{sec:conclusion}

In this study, we proposed the GG-AVSE framework to address target-speaker ambiguity in multi-talker scenarios, a critical challenge for conventional AVSE systems. To the best of our knowledge, this work is among the first to integrate gaze into modern AVSE frameworks, enabling explicit identification of the attended speaker and supplying the corresponding visual features for enhancement. Experiments on two-speaker mixtures validate its effectiveness, showing consistent and notable improvements in PESQ, STOI, and SI-SDR scores over gaze-free baselines. These findings highlight the potential of gaze as a practical auxiliary modality. Future work will focus on integrating the framework with wearable eye-tracking devices to support real-world deployment.

\vfill\pagebreak

\bibliographystyle{IEEEbib}
\bibliography{strings,refs}

\begin{thebibliography}{10}

\bibitem{cherry1953some}
E.~C. Cherry,
\newblock ``Some experiments on the recognition of speech, with one and with two ears,''
\newblock {\em Journal of the acoustical society of America}, pp. 975--979, 1953.

\bibitem{ha1}
T.~Venema,
\newblock {\em Compression for Clinicians, Chapter 7},
\newblock Thomson Delmar Learning, 2006.

\bibitem{ha2}
H.~Levitt,
\newblock ``Noise reduction in hearing aids: a review.,''
\newblock {\em Journal of Rehabilitation Research \& Development}, vol. 38, no. 1, 2001.

\bibitem{avse1}
J.-C. Hou, S.-S. Wang, Y.-H. Lai, Y.~Tsao, H.-W. Chang, and H.-M. Wang,
\newblock ``Audio-visual speech enhancement using multimodal deep convolutional neural networks,''
\newblock {\em IEEE Transactions on Emerging Topics in Computational Intelligence}, pp. 117--128, 2018.

\bibitem{avse2}
S.-Y. Chuang, H.-M. Wang, and Y.~Tsao,
\newblock ``Improved lite audio-visual speech enhancement,''
\newblock {\em IEEE/ACM Transactions on Audio, Speech, and Language Processing}, vol. 30, pp. 1345--1359, 2022.

\bibitem{avse3}
R.~Gao and K.~Grauman,
\newblock ``Visualvoice: Audio-visual speech separation with cross-modal consistency,''
\newblock in {\em Proc. CVPR}. IEEE, 2021, pp. 15490--15500.

\bibitem{avse4}
J.~Lee, S.-W. Chung, S.~Kim, H.-G. Kang, and K.~Sohn,
\newblock ``Looking into your speech: Learning cross-modal affinity for audio-visual speech separation,''
\newblock in {\em Proc. CVPR}, 2021, pp. 1336--1345.

\bibitem{michelsanti2021overview}
D.~Michelsanti, Z.-H. Tan, S.-X. Zhang, Y.~Xu, M.~Yu, D.~Yu, and J.~Jensen,
\newblock ``An overview of deep-learning-based audio-visual speech enhancement and separation,''
\newblock {\em IEEE/ACM Transactions on Audio, Speech, and Language Processing}, vol. 29, pp. 1368--1396, 2021.

\bibitem{10193049}
I.-. Chern, K.-H. Hung, Y.-T. Chen, T.~Hussain, M.~Gogate, A.~Hussain, Y.~Tsao, and J.-C. Hou,
\newblock ``Audio-visual speech enhancement and separation by utilizing multi-modal self-supervised embeddings,''
\newblock in {\em Proc. ICASSP}, 2023, pp. 1--5.

\bibitem{chao2025leveraging}
R.~Chao, W.~Ren, Y.-J. Li, K.-H. Hung, S.-F. Huang, S.-W. Fu, W.-H. Cheng, and Y.~Tsao,
\newblock ``Leveraging mamba with full-face vision for audio-visual speech enhancement,''
\newblock {\em arXiv preprint arXiv:2508.13624}, 2025.

\bibitem{gu2021efficiently}
A.~Gu, K.~Goel, and C.~R{\'e},
\newblock ``Efficiently modeling long sequences with structured state spaces,''
\newblock {\em arXiv preprint arXiv:2111.00396}, 2021.

\bibitem{redmon2016you}
J.~Redmon, S.~Divvala, R.~Girshick, and A.~Farhadi,
\newblock ``You only look once: Unified, real-time object detection,''
\newblock in {\em Proc. CVPR}, 2016, pp. 779--788.

\bibitem{deng2020retinaface}
J.~Deng, J.~Guo, E.~Ververas, I.~Kotsia, and S.~Zafeiriou,
\newblock ``Retinaface: Single-shot multi-level face localisation in the wild,''
\newblock in {\em Proc. CVPR}, 2020, pp. 5203--5212.

\bibitem{zhang2016joint}
K.~Zhang, Z.~Zhang, Z.~Li, and Y.~Qiao,
\newblock ``Joint face detection and alignment using multitask cascaded convolutional networks,''
\newblock {\em IEEE signal processing letters}, pp. 1499--1503, 2016.

\bibitem{qi2022yolo5face}
D.~Qi, W.~Tan, Q.~Yao, and J.~Liu,
\newblock ``Yolo5face: Why reinventing a face detector,''
\newblock in {\em Proc. ECCV}. Springer, 2022, pp. 228--244.

\bibitem{gu2023mamba}
A.~Gu and T.~Dao,
\newblock ``Mamba: Linear-time sequence modeling with selective state spaces,''
\newblock {\em arXiv preprint arXiv:2312.00752}, 2023.

\bibitem{padilla2025location}
D.-J.~A. Padilla, N.~L Westhausen, S.~Vivekananthan, and B.~T Meyer,
\newblock ``Location-aware target speaker extraction for hearing aids,''
\newblock in {\em Proc. Interspeech}, 2025.

\bibitem{anway2024real}
A.~R. Anway, B.~Buck, M.~Gogate, K.~Dashtipour, M.~Akeroyd, and A.~Hussain,
\newblock ``Real-time gaze-directed speech enhancement for audio-visual hearing-aids,''
\newblock in {\em Proc. Interspeech}, 2024.

\bibitem{GSG}
``Ganzin sol glasses: Wearable eye-tracking smart glasses,'' Available: \url{https://ganzin.com/en/sol-glasses-wearable-eye-tracker/}, 2025,
\newblock Official product page. Accessed: 2025-09-15.

\bibitem{yang2016wider}
S.~Yang, P.~Luo, C.-C. Loy, and X.~Tang,
\newblock ``Wider face: A face detection benchmark,''
\newblock in {\em Proc. CVPR}, 2016, pp. 5525--5533.

\bibitem{everingham2010pascal}
M.~Everingham, L.~Van~Gool, C.~K. Williams, J.~Winn, and A.~Zisserman,
\newblock ``The pascal visual object classes (voc) challenge,''
\newblock {\em International journal of computer vision}, pp. 303--338, 2010.

\bibitem{zheng2020distance}
Z.~Zheng, P.~Wang, W.~Liu, J.~Li, R.~Ye, and D.~Ren,
\newblock ``Distance-iou loss: Faster and better learning for bounding box regression,''
\newblock in {\em Proc. AAAI}, 2020, pp. 12993--13000.

\bibitem{chen2020simple}
T.~Chen, S.~Kornblith, M.~Norouzi, and G.~Hinton,
\newblock ``A simple framework for contrastive learning of visual representations,''
\newblock in {\em Proc. ICML}. PmLR, 2020, pp. 1597--1607.

\bibitem{10023284}
A.~L.~A. Blanco, C.~Valentini-Botinhao, O.~Klejch, M.~Gogate, K.~Dashtipour, A.~Hussain, and P.~Bell,
\newblock ``Avse challenge: Audio-visual speech enhancement challenge,''
\newblock in {\em Proc. SLT}, 2023, pp. 465--471.

\bibitem{afouras2018lrs3}
T.~Afouras, J.~S. Chung, and A.~Zisserman,
\newblock ``Lrs3-ted: a large-scale dataset for visual speech recognition,''
\newblock {\em arXiv preprint arXiv:1809.00496}, 2018.

\bibitem{graetzer2021clarity}
S.~Graetzer, J.~Barker, T.~J. Cox, M.~Akeroyd, J.~F. Culling, G.~Naylor, E.~Porter, R.~Viveros~Munoz, et~al.,
\newblock ``Clarity-2021 challenges: Machine learning challenges for advancing hearing aid processing,''
\newblock in {\em Proc. Interspeech}. ISCA, 2021, pp. 686--690.

\bibitem{thiemann2013diverse}
J.~Thiemann, N.~Ito, and E.~Vincent,
\newblock ``The diverse environments multi-channel acoustic noise database (demand): A database of multichannel environmental noise recordings,''
\newblock in {\em Proc. POMA}. ASA, 2013, pp. 35--81.

\bibitem{reddy2021icassp}
C.~K. Reddy, H.~Dubey, V.~Gopal, R.~Cutler, S.~Braun, H.~Gamper, R.~Aichner, and S.~Srinivasan,
\newblock ``Icassp 2021 deep noise suppression challenge,''
\newblock in {\em Proc. ICASSP}, 2021.

\bibitem{rix2001perceptual}
A.~W. Rix, J.~G. Beerends, M.~P. Hollier, and A.~P. Hekstra,
\newblock ``Perceptual evaluation of speech quality (pesq)-a new method for speech quality assessment of telephone networks and codecs,''
\newblock in {\em Proc. ICASSP}, 2001.

\bibitem{taal2010short}
C.~H. Taal, R.~C. Hendriks, R.~Heusdens, and J.~Jensen,
\newblock ``A short-time objective intelligibility measure for time-frequency weighted noisy speech,''
\newblock in {\em 2010 IEEE international conference on acoustics, speech and signal processing}. IEEE, 2010, pp. 4214--4217.

\bibitem{taal2011algorithm}
C.~H. Taal, R.~C. Hendriks, R.~Heusdens, and J.~Jensen,
\newblock ``An algorithm for intelligibility prediction of time--frequency weighted noisy speech,''
\newblock {\em IEEE Transactions on Audio, Speech, and Language Processing}, vol. 19, no. 7, pp. 2125--2136, 2011.

\bibitem{le2019sdr}
J.~Le~Roux, S.~Wisdom, H.~Erdogan, and J.~R. Hershey,
\newblock ``Sdr--half-baked or well done?,''
\newblock in {\em Proc. ICASSP}. IEEE, 2019, pp. 626--630.

\end{thebibliography}
\end{document}